\documentclass[a4paper,fleqn,usenatbib]{mn2e}

\usepackage{graphicx}	% Including figure files
\usepackage{aas_macros}
\usepackage{amsmath}	% Advanced maths commands
\usepackage{amssymb}	% Extra maths symbols
\usepackage{multicol}        % Multi-column entries in tables
\usepackage{bm}		% Bold maths symbols, including upright Greek
\usepackage{pdflscape}	% Landscape pages
\usepackage{gensymb}
\usepackage{epstopdf}

\usepackage[T1]{fontenc}
\usepackage{txfonts}

\title[Candidate Fields for EoR Imaging]{Pre-selection of the Candidate 
Fields for Deep Imaging of the Epoch of Reionization with SKA1-Low}

\author[Zheng et al.]{
Qian Zheng$^{1}$\thanks{E-mail: qzheng@shao.ac.cn,wxp@bao.ac.cn,guoquan@shao.ac.cn},
Xiang-Ping Wu$^{1, 2}$ ,
Quan Guo$^{1}$,
Melanie Johnston-Hollitt$^{3,4}$,
\newauthor Huanyuan Shan$^{1,5}$,
Stefan W. Duchesne$^{4}$,
Weitian Li$^{6}$
\\
$^{1}$Shanghai Astronomical Observatory, Chinese Academy of Sciences\\ 
  80 Nandan Road, Shanghai 200030, China\\
$^{2}$National Astronomical Observatories, Chinese Academy of Sciences\\ 
  20A Datun Road, Beijing 100012, China\\
$^{3}$Curtin Institute for Computation, GPO Box U1987, Perth, WA 6845, Australia\\ 
$^{4}$International Centre for Radio Astronomy Research (ICRAR), Curtin University, Bentley, WA, Australia\\
$^{5}$University of Chinese Academy of Sciences, Beijing 100049, China \\  
$^{6}$School of Physics and Astronomy, Shanghai Jiao Tong University\\
  800 Dongchuan Road, Shanghai 200240, China\\
}

\date{Last updated 2018 January 31; in original form 2017 XXX X}

\pubyear{2015}

\begin{document}
\label{firstpage}
\pagerange{\pageref{firstpage}--\pageref{lastpage}}
\maketitle

\begin{abstract}
The Square Kilometre Array (SKA) will be the first low-frequency instrument with the capability to directly image the structures of the Epoch of Reionization (EoR). Indeed, deep imaging of the EoR over 5 targeted fields of 20 square degrees each has been selected as the highest priority science objective for SKA1. Aiming at preparing for this highly challenging observation, we perform an extensive pre-selection of the `quietest' and `cleanest' candidate fields in the southern sky to be suited for deep imaging of the EoR using existing catalogs and observations over a broad frequency range. The candidate fields should meet a number of strict criteria to avoid contaminations from foreground structures and sources. The candidate fields should also exhibit both the lowest average surface brightness and smallest variance to ensure uniformity and high quality deep imaging over the fields. Our selection eventually yields a sample of 7 `ideal' fields of 20 square degrees in the southern sky that could be targeted for deep imaging of the EoR. Finally, these selected fields are convolved with the synthesized beam of SKA1-low stations to ensure that the effect of sidelobes from the far field bright sources is also weak. 
\end{abstract}


%%%%%%%%%%%%%%%%%%%%%%%%%%%%%%%%%%%%%%%%%%%%%%%%%%

%%%%%%%%%%%%%%%%% BODY OF PAPER %%%%%%%%%%%%%%%%%%

% The MNRAS class isn't designed to include a table of contents, but for this document one is useful.
% I therefore have to do some kludging to make it work without masses of blank space.

\section{Introduction}
\label{sec:intro}
Exploration of the dark ages, cosmic dawn (CD) and epoch of reionization (EoR) constitutes the last frontier of observational cosmology for the next decades, which will unveil the mysteries of how and when the universe underwent a transition from a dark to bright phase, how the underlying large-scale structures grew from linear to nonlinear stage, and how baryonic matter became prominent for formation of cosmic structures. It has also been recognized that a direct probe of the 21 cm radiation associated with the neutral hydrogen beyond redshifts $z>6$ would make this observational campaign possible, as the neutral hydrogen, amounting to 75\% of the baryonic mass of the universe, can be considered as a good tracer of the properties of formation and evolution of the universe in its early stage (for reviews see \citealp{Furlanetto06}; \citealp{pritchard12}; \citealp{zaroubi13}; \citealp{ferrara14}). Essentially, there are three observational approaches to directly measuring the CD/EoR: (1) the spatially averaged global signature at each frequency, (2) the spatial fluctuations via statistical properties revealed by power spectra, and (3) tomographic imaging of the ionized structures. While many dedicated low-frequency radio facilities have been built or planned around the globe to search for either the mean brightness over all directions in the frequency range of 50-200 MHz [e.g. EDGES \citep{Bowman10,Bowman08} with recent detection of an absorption signature at about redshift 17 \citep{Bowman18}, BIGHORNS \citep{Sokolowski15}, SCI-HI \citep{Voytek14}, LEDA \citep{Bernardi15}, SARAS \citep{Patra13}, PRI$^Z$M \citep{Philip19}. etc.] or the statistical measurement of fluctuations in the redshifted 21 cm backgrounds [e.g. 21CMA \citep{Zheng16}, LOFAR \citep{vH13,Greig20}, LWA \citep{Taylor12}, MITEoR \citep{zheng14}, MWA \citep{Bowman13,Tingay13}, PAPER \citep{Jacobs11}, GMRT \citep{Paciga13}, HERA \citep{deboer17}, etc.], the SKA-Low will be the first low-frequency instrument with the capability of directly imaging the structures of CD/EoR - probably the most exciting scientific outcome of the SKA-Low.

Imaging of CD/EoR was ranked first among the thirteen highest priority science objective selected by the Science Review Panel of SKA-Low in December 2014 \citep{Koopmans15}. In terms of current design and observational strategy, deep observations of 5000 hours integration time in total over 5 targeted-area of 20 square degrees each will be performed with SKA1-Low (\citealp{Mellema15}; \citealp{Koopmans15}). This will provide $\sim\!1$~mK image noise levels under 0.1 MHz spectral resolution and on angular scales of a few to a few tens of arcminutes, depending on observing frequencies. Yet, the imaging of EoR below 100 MHz seems unfeasible due to the loss of sensitivity with current SKA1-Low baseline design, although some of the giant isolated `ionized' regimes of CD/EoR can still be mapped out at $\sim\!10$~mK noise level with somewhat poor angular resolutions ($\sim\!10'$). Therefore the statistical measurements of the power spectrum will be the only way to probe the structures of CD and early EoR stages beyond redshift $\sim\!13$. With much larger collecting area, better angular resolution and wider frequency coverage, SKA2 should eventually enable us to reach our ultimate goals. Another key parameter for imaging the EoR is the adequate field-of-view (FoV), which should be at least a few degrees, in order to image the largest ionized structures during the later EoR stages. With the design of SKA1-Low deployment baselines, the primary beam of the 38 m station varies from $6^{\circ}$ at 100 MHz to $3^{\circ}$ at 200 MHz, which should meet the purpose. 

The deep 1000-hr integrations for each of the 5 separated EoR fields with SKA1-Low are only the minimum requirement to reach the sensitivity of imaging the ionized structures. A considerably large number of radio quiet nights, adding up to 3--5 years, may be needed if the SKA1-low operates in drift-scan mode. The tomographic imaging reconstruction for SKA1-Low as an interferometer at low frequencies suffers from various systematic errors, including radio frequency interference, incomplete sampling, limited angular resolution, ionospheric distortion, inaccurate calibration, instrumental artifacts, poor sky modeling, confusion limit, etc. A small error from incomplete corrections to any of these factors may destroy the imaging recovery of the EoR structures which are deeply buried under both astronomical foregrounds ($\sim\,$5 orders of magnitude stronger than the EoR signal) and instrumental noises. Many state-of-the-art techniques and algorithms have thus been developed in past decades aiming to overcome these observational and technical hurdles, which may hopefully allow one to reach the desired detection sensitivity ($\sim\!1$~mK) for the future 1000 hours integrated observations of the EoR field with SKA1-Low. 

Aiming at preparing for the deep imaging of the EoR ionized structures with SKA1-Low which will be fully operated in 2028, it is timely to select the five (or more) candidate fields of 20 square degrees each in the southern sky to be suited for this purpose. Pre-observations and preliminary searches with existing radio facilities such as MWA, ASKAP, and even the Karl G. Jansky VLA and FAST (if the fields are observable in the northern hemisphere) over a wide waveband should be made in order to understand the `quietness' and `cleanness' of the candidate fields and their environmental effects. This will help to answer the questions of how foreground bright cosmic sources, especially the extended and diffuse ones, are distributed and clustered in the fields, to what extent the sky models can be constructed and foreground sources removed, how sidelobe leakages of off-field bright sources affect the imaging quality and sensitivity, and what kinds of noise remain to be the major source of errors in imaging reconstruction and what kinds of technologies should be needed and further developed to beat down the noise(s) to an acceptable level. After all these exercises we will hopefully be able to determine whether the candidate fields can eventually be selected for the first EoR imaging observations before the first light of the SKA1-Low. With such motivation in mind, we attempt to fulfill the task in the present work by making a pre-selection of the candidate EoR fields, based mainly on the existing radio catalogs and observations.

\section{Target Selection}
\label{sec:targetselection}
\subsection{Locations and Sizes of the Candidate Fields} \label{subsec:location}
SKA1-Low will be deployed at the remote site close to the Murchison Radio astronomy Observatory (MRO) in Western Australia. The telescope receptors of SKA1-Low will consist of 131,072 log-periodic dual-polarized antennas grouped into 512 stations of 38 m in diameter, operating in the frequency range of 50 MHz -- 350 MHz. While SKA1-Low is designed to survey a significant fraction of the sky up to 10000 square degrees, its unprecedentedly high sensitivity allows one to detect extremely faint objects and study their structures. Within a long integration of 1000 hours, it is expected that SKA1-low can reach $\sim\!1$~mK noise level under 0.1 MHz spectral resolution at frequencies of $\nu>100$ MHz. SKA1-low will have a sky coverage similar to that of MWA, which can observe the entire sky south of Dec. +30$^\circ$ \citep{Hurleywalker17}. For this reason, the first criterion for selecting the EoR candidate fields is to restrict the fields to the sky south of Dec. $+30^\circ$. In reality, priority will be given to the fields with smaller zenith angles (typically $<50^\circ$) to avoid the non-uniform beam pattern.

\begin{figure*}
\begin{center}
\vspace{-0.3cm}
%\vspace{2cm}
 \includegraphics[width=16.0cm]{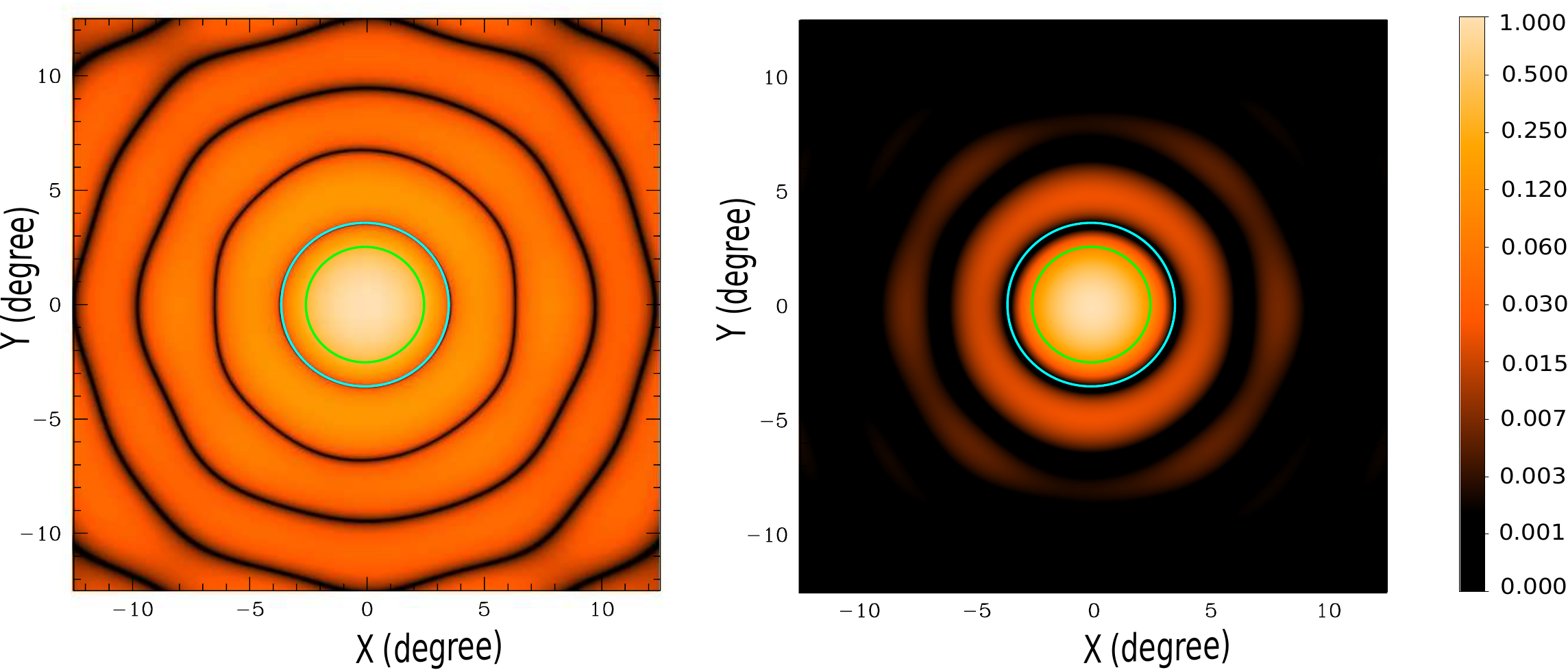}
%\vspace{-0.4cm}
 \caption{The simulated beam pattern of one SKA1-low station (left) and the synthesized beam pattern of two stations (right) toward zenith at 150 MHz. The fields of 20 and 40 square degrees are plotted in green and blue lines, respectively.}
\label{fig:skabeam}
\end{center}
\end{figure*}

\begin{figure*}
\begin{center}
\vspace{-0.1cm}
%\vspace{2cm}
 \includegraphics[width=18.0cm]{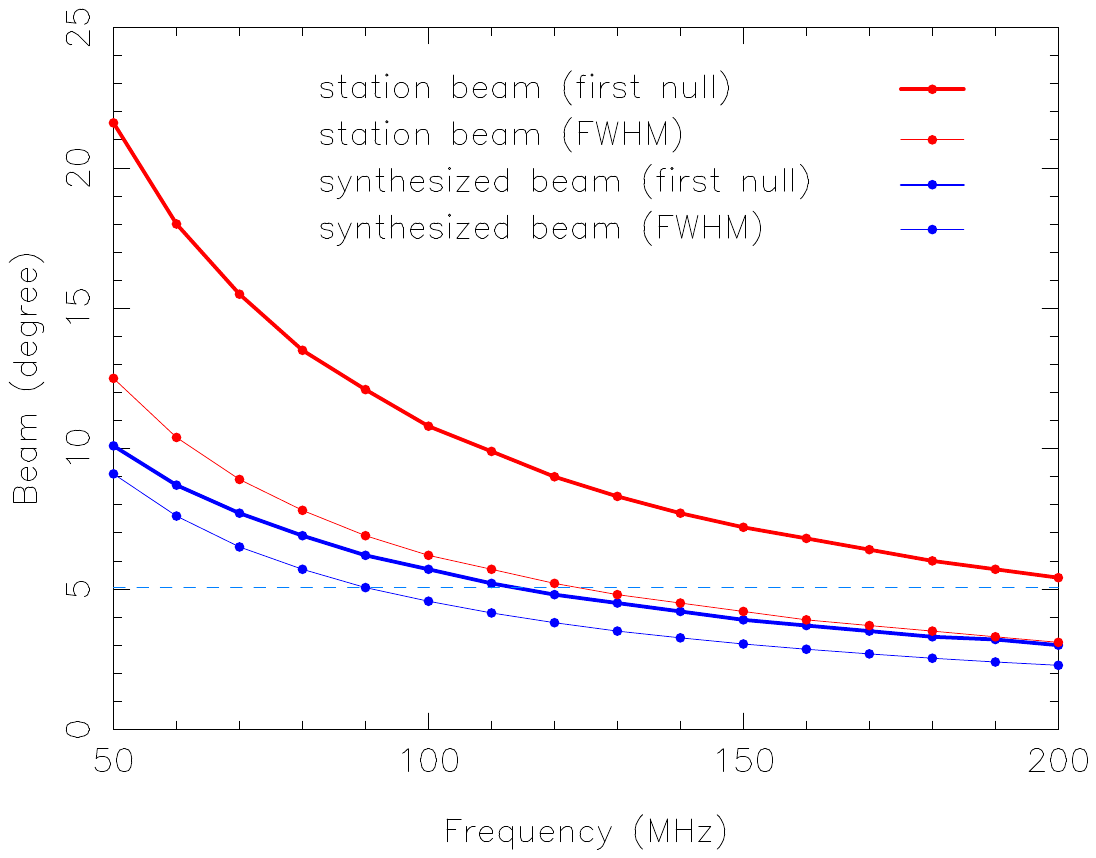}
%\vspace{-0.4cm}
 \caption{An example of the beam pattern of a single SKA1-low station and the synthesized beam of two SKA1-low stations at different frequencies. The primary beams at both first null and FWHM are shown. The dashed light-blue line indicates the size of 20-square-degrees for deep imaging of EoR.}
\label{fig:stations}
\end{center}
\end{figure*}

The candidate fields should also be large enough to capture the largest ionized structures ($\sim 100$ Mpc) during the very late stage of EoR near 200 MHz or $z=6$ (e.g. \citealp{zaroubi13,Geil17}). We have therefore simulated the beam patterns of two SKA1-low stations based on randomly positioned 512 SKALA4 elements each in a circle of 38 m in diameter in terms of the updated design of SKA1 (\citealp{Dewdney19,Acedo20,Acedo20b}). Figure \ref{fig:skabeam} shows an example of the beam pattern for a single SKA1-low station and also the synthesized beam of two stations toward the zenith. For the latter a baseline of 5 km is assumed. In Figure \ref{fig:stations} we plot the frequency dependence of the station beam and synthesized beam at both first null and FWHM. A comoving scale of 100 $h^{-1}$ Mpc at $z=6$ corresponds approximately to an angular scale of roughly $1^{\circ}$ $h$. So, the current design of SKA1-low with a primary beam of $3^{\circ}$ (FWHM) and a synthesized beam of $2^{\circ}$ (FWHM)  at 200 MHz should enable us to observe the entire ionized structures even near the end of the EoR, if other observational constraints such as sensitivity, resolution and foreground subtraction are left aside. The synthesized beam of SKA1-low would reach $\sim5^{\circ}$ (FWHM) and $\sim 6^{\circ}$ (first null) at 100 MHz, probably the lowest frequency at which we can capture the EoR imaging with SKA1-low due to rapidly decreasing sensitivity with decreasing frequency. This defines the field of view of 20 square degrees (2.52 degrees in radius) for the deep image of the EoR structures in the frequency range of 100-200 MHz with SKA1-low. So, our second criterion for selecting the EoR candidate fields is to choose an adequate FoV that should be larger than $2.52^{\circ}$ in radius. Nonetheless, for comparison we also adopt a relaxed condition of $3.57^{\circ}$ in radius (corresponding to a FOV of 40 square degrees) to see whether we could find even larger radio quiet areas to suppress possible environmental effects.

\subsection{Galactic Emission} \label{subsec:galactic}

\begin{figure*}
\begin{center}
\vspace{-0.3cm}
%\vspace{2cm}
 \includegraphics[width=13.0cm]{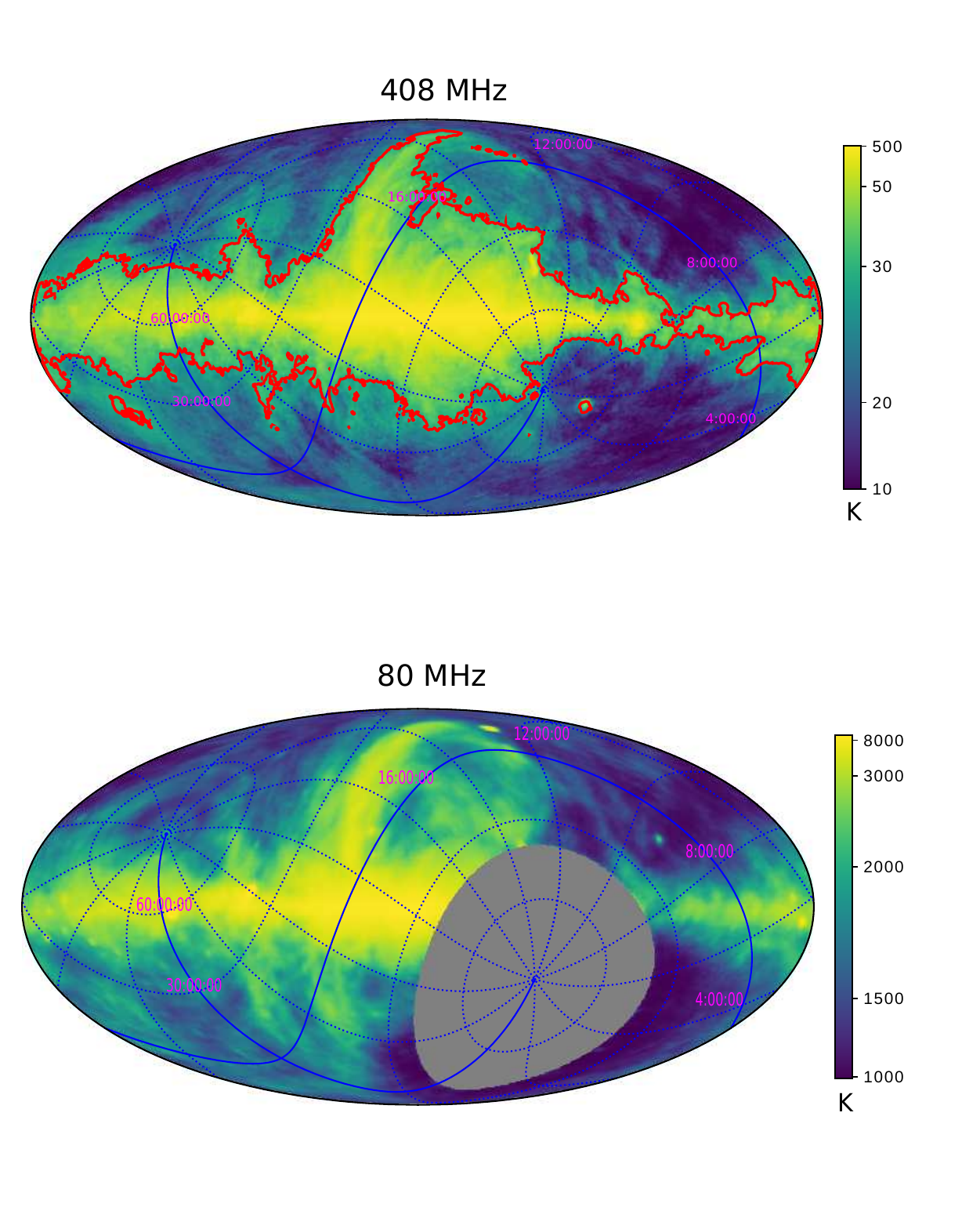}
%\vspace{-0.4cm}
 \caption{The Haslam 408 MHz All-Sky Map (top) and 80 MHz LWA map (bottom) in Galactic coordinates. The red contours on the Haslam 408 MHz All-Sky Map indicate the 30 K level, and the EoR candidate fields will be selected in the regions with temperature below this limit.}
\label{fig:milkyway}
\end{center}
\end{figure*}

The low-frequency radio sky is dominated by diffuse synchrotron emission from our Galaxy, which is on average 4-5 orders of magnitudes brighter than the EoR surface brightness at frequencies 50--200 MHz. Unfortunately, below 200 MHz little is known about the spatial and spectral properties of diffuse Galactic emission despite that many investigations have been made thus far (see, e.g. \citealp{deOliveiraCosta}; \citealp{zheng17}). Like many previous studies of modeling of the low-frequency emission of the Milky Way, our selection of the EoR candidate fields begins with the Haslam 408 MHz All-Sky Map \citep{Haslam70,remazeilles15}, which allows us to identify the `cooler' regions in the southern sky. In principle, by extrapolating the Haslam 408 MHz All-Sky Map with a proper assumption of the spectral index, one can build the global sky model at the EoR window, 50--200 MHz \citep{deOliveiraCosta,zheng17}. We further use the recent 80 MHz LWA map \citep{dowell17}, which covers the sky north of a declination of $-40^{\circ}$, to cross-check these `cooler' or radio-quiet regions. Figure~\ref{fig:milkyway} shows the Haslam 408 MHz All-Sky Map and the 80 MHz LWA map, respectively. The regions with temperature below 30 K in the Haslam 408 MHz All-Sky Map are defined as 'cooler' ones in this work, which will be the targeted fields for our selection of the EoR candidate fields.

\subsection{The Magellanic Clouds}\label{subsec:diffuse}

\begin{figure*}
\begin{center}
\vspace{-0.3cm}
%\vspace{2cm}
 \includegraphics[width=13.0cm]{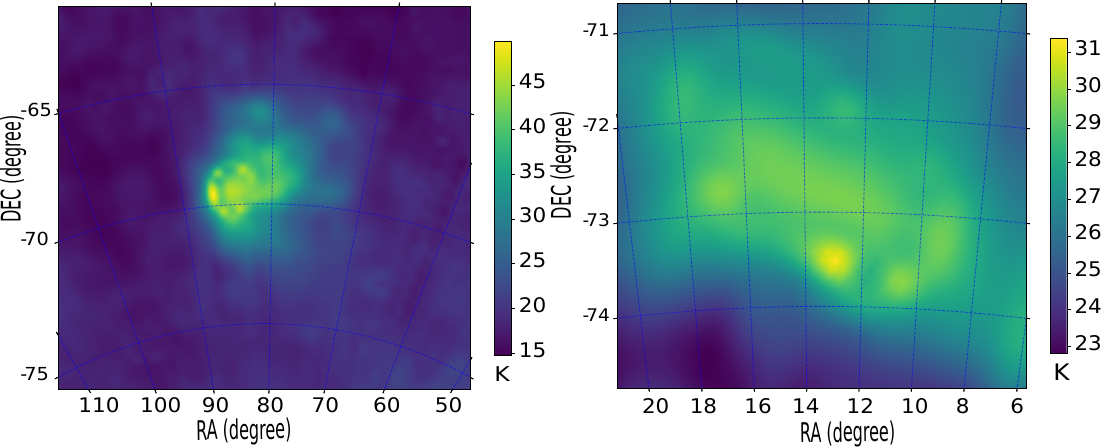}
%\vspace{-0.4cm}
 \caption{The Large Magellanic Cloud (left) and Small Magellanic Cloud (right) in the Haslam 408 MHz All-Sky Map.}
\label{fig:LMC}
\end{center}
\end{figure*}

Besides the Galactic plane, the Large and Small Magellanic Clouds (LMC/SMC) are the most prominent extended celestial sources in the southern sky (Figure~\ref{fig:LMC}). In order to avoid both the possible obscuration/occultation of observations of the cosmic EoR structures and the difficulty of subtracting complex foregrounds, our candidate fields should be chosen far away from the Magellanic Clouds. 
%------------------------------------------------------------
% Figure  : 
%------------------------------------------------------------

\subsection{Bright Radio Sources}\label{subsec:bright}
Point sources at low frequencies spread over the whole sky, and constitute one of the biggest challenges in detection of the EoR signal in addition to the contamination of the Galactic emission. Therefore, there are primarily three rules that we should follow in selecting the candidate EoR fields: (1)The EoR fields of 20 square degrees should not contain any bright radio sources; (2)The EoR fields should look somewhat `isolated', i.e., there are as few bright radio sources as possible in both the vicinities of the fields and the sidelobes of the SKA1-low stations. Therefore, we will also survey an area of twice big as the EoR imaging fields for comparison, namely 40 square degrees or $3.57^{\circ}$ in radius; (3)Since the EoR fields inevitably contain foreground faint radio sources, we need to select the fields over which the faint radio sources distribute as uniformly as possible. The last point is to ensure that the EoR imaging would have a uniform noise level across the entire field after the CLEAN processing and foreground source subtraction. Clustering of the radio sources in the fields may add difficulty to these works.

The bright radio sources may refer to those with flux densities above a few Jy. Using a broken power-law for the differential radio source counts \citep{Hales88,Cohen07,Moore13} and assuming a mean spectral index of $-0.8$, we can estimate how many foreground radio sources would be encountered at different fluxes over a field of up to 40 square degrees. It turns out that at frequency of 150 MHz, the surface number density of radio sources reaches roughly $\sim0.05$ and $\sim0.1$ per square degree at 5 and 3 Jy, respectively. On average a field of 40 square degrees contains roughly $\sim 2$ ($\sim 4$) radio sources with fluxes brighter than 5 (3) Jy. Therefore, it should be possible to find the EoR candidate fields of 40 square degrees without containing any bright point sources with flux of greater than a few Jy at 150 MHz. If we restrict the EoR field to 20 square degrees, we could even choose a lower flux threshold of $\sim 1$ Jy. Yet, in either case, we should also need efficient algorithms to remove all the faint sources below the flux threshold.

Two catalogs are employed to search for the bright sources: the public release of the 150 MHz continuum survey TGSS alternate data release 1(TGSS ADR1 \citep{intema16}) from GMRT, which covers the full sky north of DEC -53 degrees with a median RMS background noise of 3.5mJy per beam and a typical resolution of 25 arcseconds \citep{intema16}, and the GaLactic and Extragalactic All-sky MWA (GLEAM) survey \citep{Wayth15}, which observed half of the sky up to 24,831 square degrees over declinations south of $+30^\circ$ and Galactic latitudes outside $10^\circ$ of the Galactic plane - almost the same sky coverage as that of future SKA1-low. Our search is mainly based on the TGSS catalog because of its higher resolution, but for the southern sky of declinations $\delta\leq-53^\circ$, the GLEAM catalog is used. It should also mention that differential source counts from the GLEAM extragalactic catalog \citep{Hurleywalker17} yield a similar bright source surface number density as the above estimates.

Defining each of the EoR candidate fields as an circular area of 20 square degrees, we perform a survey of the radio `quiet' fields over the TGSS/GLEAM sky which do no contain bright radio sources of $S\geq1,2,3,4,5,6$ Jy at 150 MHz. This yields a total of (6,128, 266, 342, 366, 369) circular fields for $S\geq1,2,3,4,5,6$ Jy, respectively, over the TGSS/GLEAM sky. These numbers decrease to (1, 14, 50, 115, 147, 193) if we adopt a larger circular area of 40 square degrees.

\begin{figure*}
\begin{center}
\vspace{-0.3cm}
%\vspace{2cm}
\includegraphics[width=12.0cm]{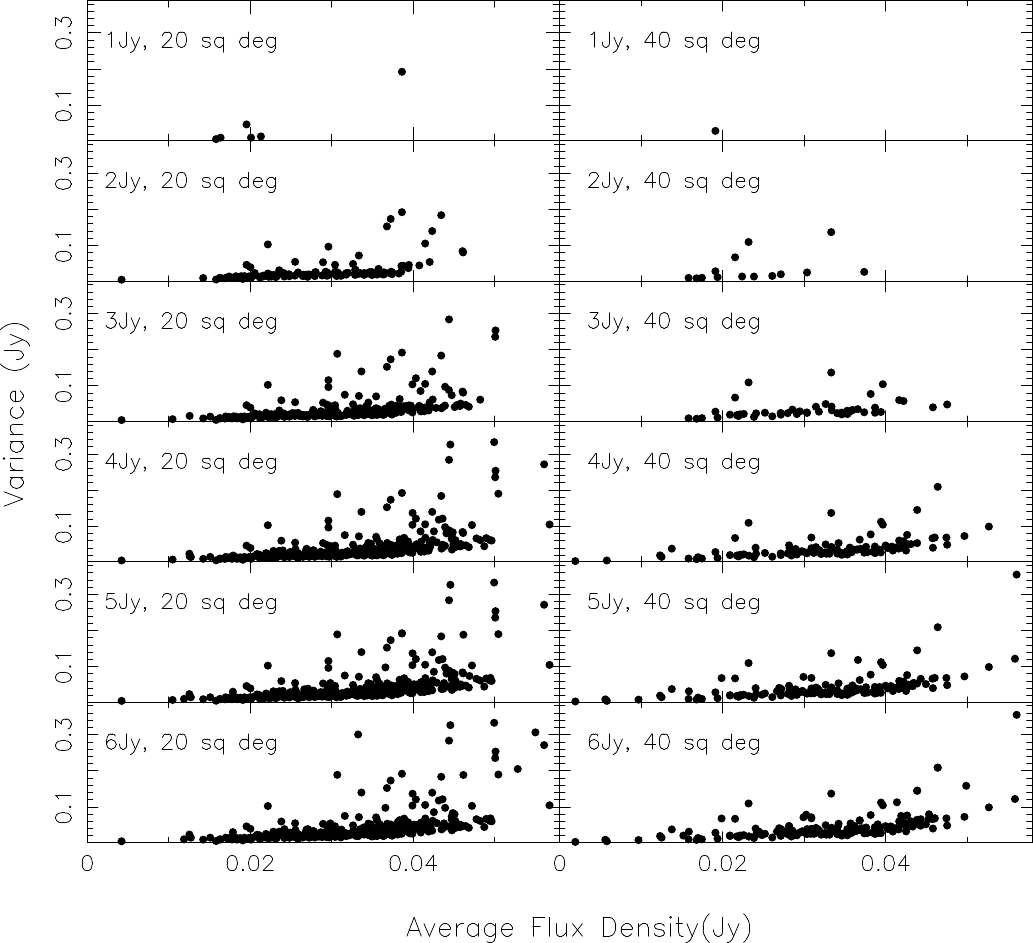}
%\vspace{-0.4cm}
\caption{The mean flux against flux variance for all the candidate fields selected from the TGSS/GLEAM survey. Six flux thresholds (1, 2, 3, 4, 5 and 6 Jy) and two field limits (20 and 40 square degrees) are adopted. }
\label{fig:variance-bright}
\end{center}
\end{figure*}

Note that not all the selected fields are actually radio `quiet' and hence suited for deep imaging of EoR structures. Some of them may contain too many faint sources, which as a whole may look rather bright, and/or the faint sources may exhibit clumpy or clustering which would add extra work or difficulties for CLEAN or subtraction processes. To this end, we calculate the mean flux and variance of each selected field and plot the results in Figure \ref{fig:variance-bright}. While the mean fluxes for all the fields are smaller than $\sim0.05$ Jy, a few of the selected fields demonstrate rather large flux variance especially in the case of a smaller field of 20 square degrees. As a comparison, the average sky flux and variance are $0.52 \pm 0.13$ Jy and $1.51 \pm 1.09$ Jy, respectively. Our selected fields are typically one order of magnitude fainter in flux and smoother in spatial distribution than the averages. We will come back to this point later when we sort out the final list of candidate fields by applying and combining all the criteria.

%------------------------------------------------------------
% Figure  : 
%------------------------------------------------------------

\subsection{Diffuse Sources and Galaxy Clusters}\label{subsec:cluster}
Foreground diffuse and extended radio sources in low-frequency range could mimic the EoR structures and constitute one of the biggest problems for recovery of the true EoR signals from deep imaging observations \citep{Jelic08,Brunetti14,Li19}. Though these diffuse sources may be subtracted based on their smooth spectral features, a tiny error from the imperfect modeling of the extended sources may invalidate our efforts. Therefore, we will add another constraint to the EoR candidate fields: They should not contain any resolved radio diffuse sources in addition to the very extended, bright sources of the Milky Way and LMC/SMC. 

\begin{figure*}
\begin{center}
\vspace{-0.3cm}
%\vspace{2cm}
 \includegraphics[width=13.0cm]{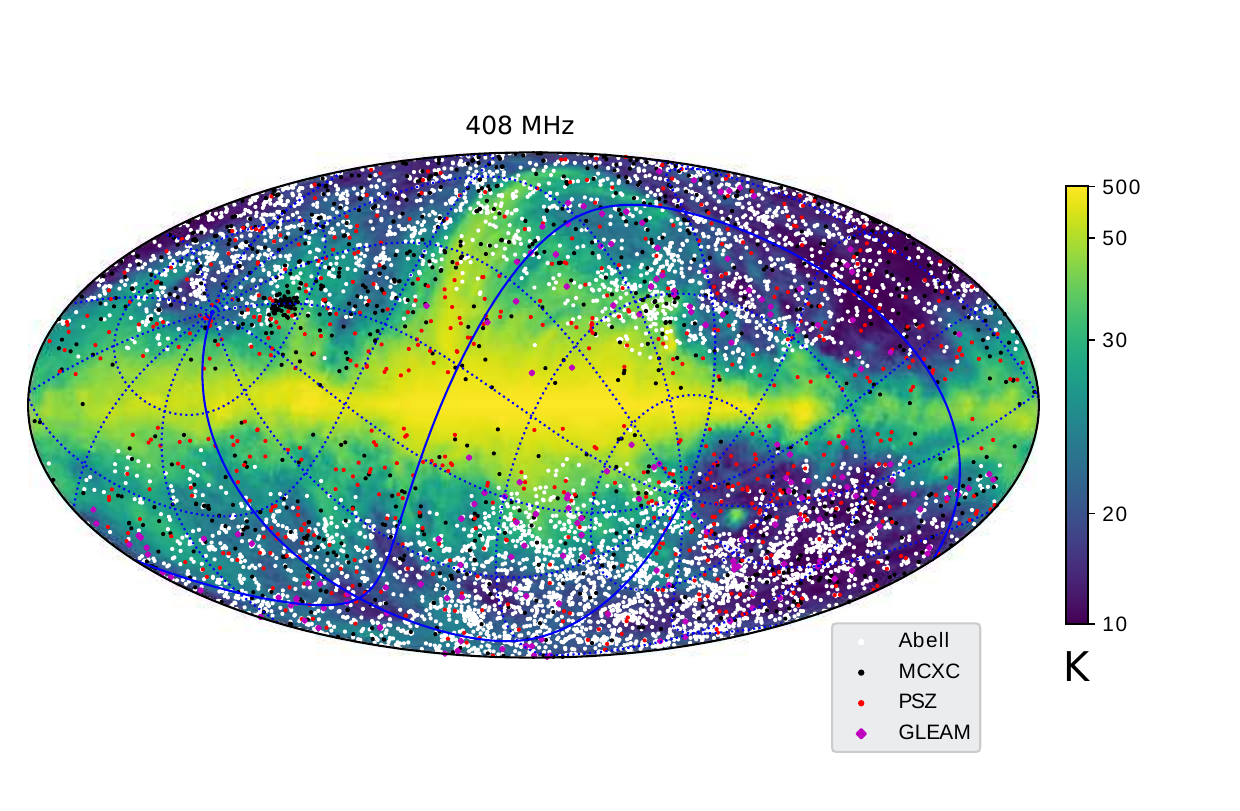}
%\vspace{-0.4cm}
 \caption{The distribution of 6986 galaxy clusters on the Haslam 408 MHz All-Sky Map: 5249 Abell clusters (Abell), 1058 X-Ray clusters (MCXC), 679 Planck SZ clusters (PSZ) and 226 galaxy clusters with diffuse emission observed with GLEAM (GLEAM). Note that many of the X-ray/SZ selected clusters and the GLEAM clusters are also the Abell clusters. }
\label{fig:cluster}
\end{center}
\end{figure*}

Diffuse radio sources in the Galaxy such as the supernova remnants can be identified and hence easily avoided. We will focus on the extragalactic extended radio sources, mainly clusters of galaxies, and many if not all of them contain diffuse radio relics and/or halos. We have therefore compiled a catalog of 6986 clusters, based on multi-wavelength observational data available in literature, which include 5249 optical Abell clusters \citep{Abell89}, 1058 X-ray clusters \citep{Piffaretti11}, 679 Planck SZ clusters and 226 clusters detected with diffuse emission by GLEAM at very low-frequencies (Johnston-Hollitt et al. in prep.). Yet, it is noted that many of the X-ray/SZ selected clusters have already been included in the Abell cluster catalog. Figure \ref{fig:cluster} shows the distribution of all these clusters over the Haslam 408 MHz All-Sky Map. The EoR candidate fields to be selected should contain no clusters if possible. 

%------------------------------------------------------------
% Figure  : 
%------------------------------------------------------------

\subsection{Effect of Sidelobes}

Although the candidate fields can be selected without any bright radio sources, it is still possible that significant power or noise is introduced by bright sources outside the fields through the sidelobes of the telescope \citep{Vedantham12,morales12,Parsons12}. Since the primary beam of an SKA1-low station varies from $5^{\circ}$ degree to $11^{\circ}$ (measured as center to first null) in the 100 - 200 MHz range, the bright sources, especially the so-called A-term sources, at tens of degrees far from the centers of the selected candidate fields could contaminate deep observations. So, as a final step, we have to quantitatively estimate the effect of the sidelobe for each of the candidate fields, using the beam model of the SKA1-low, and make sure that the sidelobe noise is lower than thermal noise (typically $\sim 1$ mK). To this end, all the sources within 20 degrees around each of the candidate fields will be taken into account. Also, we decide not to include the  fields in our final list if no observational data from TGSS/GLEAM are available over their surrounding areas up to 20 degrees. Furthermore, the effect will be demonstrated with an imaginary observation at noon on January 27, 2025 at the location of the MRO (longitude 116.7$^{\circ}$, latitude 26.7$^{\circ}$).

\section{Results}
We now summarize the criteria for selecting the candidate fields for
deep imaging of EoR with the future SKA1-low:\\
(1) located within the cold regions with temperature below 30 K 
    in the Haslam 408 MHz All-Sky Map;\\
(2) far away from the LMC, SMC and other bright, extended structures; \\
(3) containing no bright radio point sources;\\
(4) containing no massive clusters of galaxies;\\
(5) exhibiting both lowest mean surface brightness and lowest surface brightness variance;\\
(6) selecting only one field when the overlapping area of two adjacent candidate fields 
exceeds $20\%$;\\
(7) sidelobe contaminations well below $\sim1$ mK\\

Although our priority choice of FoV is 20 square degrees or $2^{\circ}.52$ in radius, the size that will be adopted for the deep imaging of EoR with SKA1-low in terms of current observational strategy, we also provide the candidates for a larger field of 40 square degrees or $3.57^{\circ}$ in radius for both comparison and optional choices. The key difference between the small (20 square degrees) and large (40 square degrees) fields is the flux threshold $S_{\rm lim}$ that is used to define the so-called bright radio sources (see section 2.4). Using a higher cut of $S_{\rm lim}$ would ensure a sufficiently large number of candidate fields (e.g. $\ge5$) to be selected, though the noise level would simultaneously increases. 

\begin{figure*}
\begin{center}
\vspace{-0.3cm}
%\vspace{2cm}
 \includegraphics[width=12.0cm]{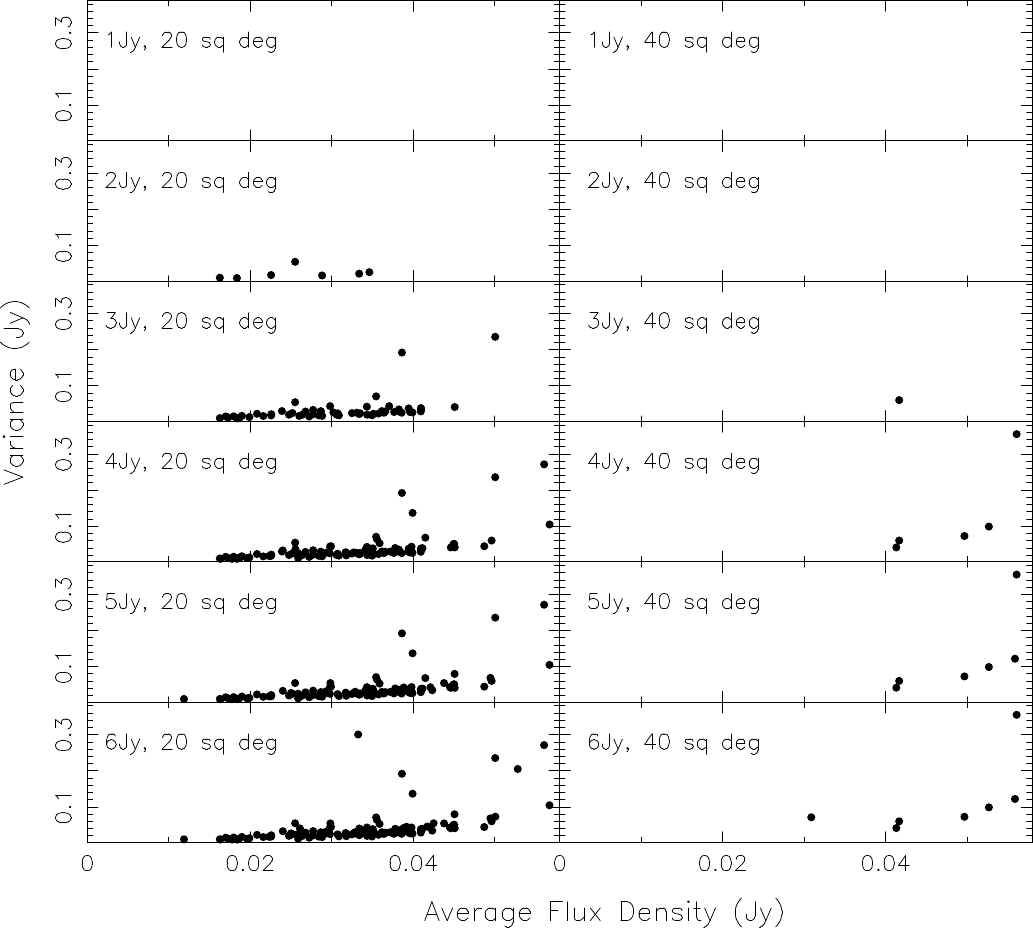}
%\vspace{-0.4cm}
 \caption{The average flux against flux variance for each of the candidate 
fields selected in terms of different flux thresholds for bright radio sources.  
Two choices of field of view, 20 and 40 square degrees, are used.}
\label{fig:variance-final}
\end{center}
\end{figure*}

\begin{figure*}
\begin{center}
  \vspace{-0.3cm}
  \includegraphics[width=14.0cm]{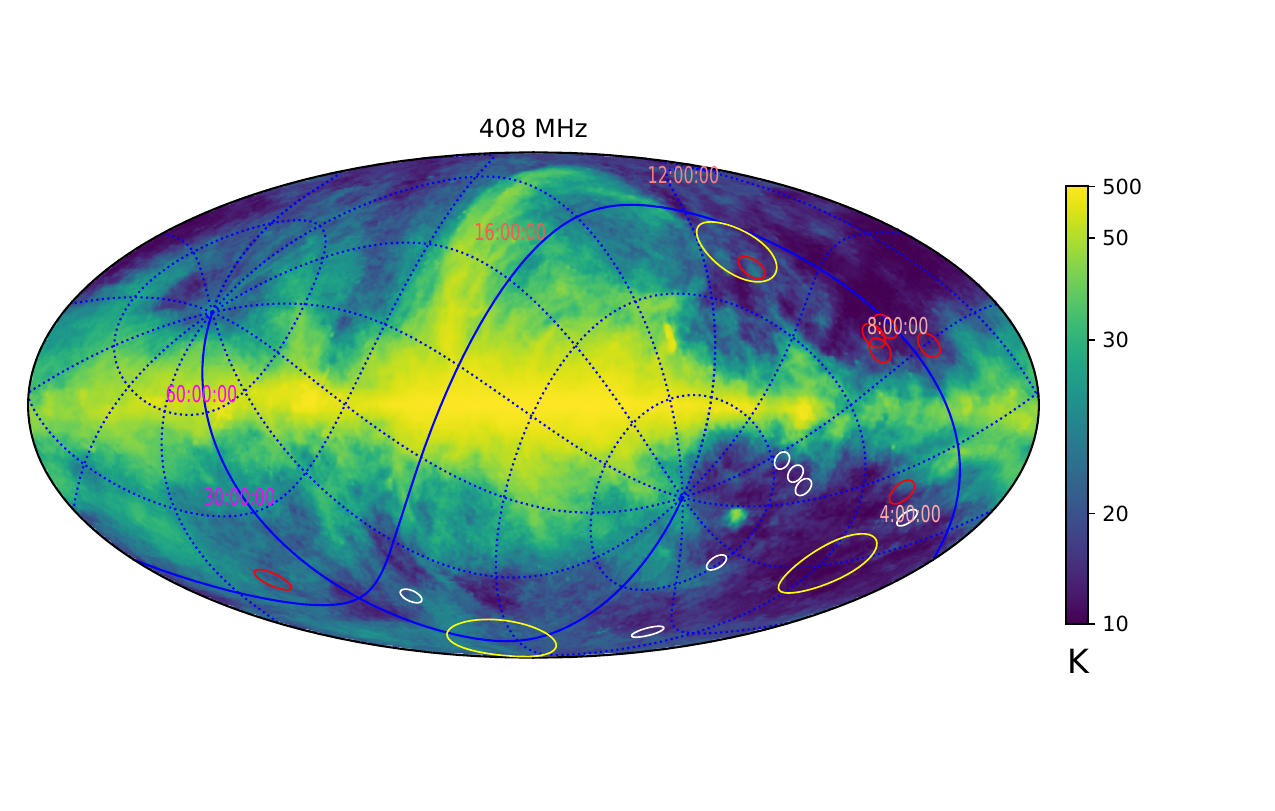}
  \caption{Locations of seven 20-square-degree potential candidate fields (white circles) and seven 40-square-degree potential candidate fields (red circles)  in the Haslam 408 MHz All-Sky Map. The three MWA EoR fields (yello circles) are also displayed for comparison.}
%\vspace{2cm} 
%\vspace{-0.4cm}
 \label{fig:regions}
\end{center}
\end{figure*}

Keeping two choices of field sizes (20 and 40 square degrees), we now apply the above selection criteria to the southern sky by varying the flux threshold of bright sources $S_{\rm lim}$. Figure \ref{fig:variance-final} displays the total number of candidate fields, their average fluxes and variance for different values of $S_{\rm lim}$. It appears that no candidate field can be found if flux threshold is taken to be lower than 1 Jy and 2 Jy for the 20 and 40 square degree cases, respectively. For a larger field size of 40 square degrees, we can find five candidate fields with a bright sources limit of $S_{\rm lim}\leq4$ Jy at 150 MHz. Seven candidates are reached if we raise the flux limit to $S_{\rm lim}\leq6$ Jy at the same frequency, although one of the fields demonstrates a larger flux variance. For a choice of smaller field of $20$ square degrees, seven candidate fields can also be found at $S_{\rm lim}\leq2$ Jy at 150 MHz. Eventually, we decide to take seven candidate fields each as our final samples for the 20 and 40 square degree cases, and their properties (coordinates, number of radio sources and the brightest one) are summarized in Table \ref{table:fields20} and Table \ref{table:fields40}, respectively. Locations of these potential candidate fields are displayed in Figure \ref{fig:regions}. Overall, the small field sample is better than the large field one in terms of radio quietness. 

One of the 40-square-degree candidate fields is located within one of the three MWA EoR fields (see Figure \ref{fig:regions}) (\citealp{Pindor18,Procopio17,Barry19}). The latters are selected to avoid both the brightest regions of the Milky Way and the bright extragalactic radio sources while maximizing the observing time over the year. Nonetheless, the brightest radio sources in all the three MWA EoR fields ($\sim 20^{\circ}\times20^{\circ}$) are approximately 20 Jy at 150 MHz, an order of magnitude brighter than the ones in our candidate fields.  Similarly, one of the LOFAR EoR fields is even chosen to contain very bright radio source 3C169 for the purpose of calibration and understanding of the systematics \citep{deBruyn12,Yatawatta13}. Indeed, the EoR fields with current SKA-low pathfider/precusor are targeted at statistical study of the EoR signal, and therefore are often ten times larger than our candidate fields dedicated to direct imaging of the EoR structures. Under this circumstance it is almost impossible to avoid including the bright extragalactic radio sources of  $S>10$ Jy. The lack of very bright radio sources in our candidate fields does not constitute a major problem for absloute calbration because there are still a few hundreds of relatively uniformaly distributed radio point sources that can be used for the purpose.

%------------------------------------------------------------
% Figure  : 
%------------------------------------------------------------

%----0.609857---0.2,0.05,0.01,0.001--------------------------------------------------------------
\begin{table*}
%\onecolumn
  \centering
  \caption{Seven candidate fields of 20 square degrees.}
\begin{tabular}{c|c|c|cc|}
\hline
  field  & R.A.(J2000),Dec.(J2000) & flux threshold & 
number of sources & brightest source \\
& (h:m:s, d:m:s) & (Jy) &  & (Jy) \\
\hline
1 &(1:39:12,-39:42:00) & 2 & 190  & 2.00 \\
2 &(3:23:36,-60:00:00) & 2 & 162  & 1.93 \\
3 &(4:48:00,-13:24:00) & 2 & 357  & 1.52 \\         
4 &(6:28:24,-49:42:00) & 2 & 215  & 1.94 \\
5 &(7:00:48,-52:24:00) & 2 & 176  & 1.83 \\
6 &(7:41:12,-56:24:00) & 2 & 182  & 1.55 \\
7 &(23:10:00,-8:06:00) & 2 & 331  & 1.84 \\
\hline
\end{tabular}
\label{table:fields20}
\end{table*}
%---------------------------------------------------------------------
% <3,4,5,6 Jy, 40 square degrees
%---------------------------------------------------------------------
\begin{table*}
%\onecolumnop00
  \centering
  \caption{Seven candidate fields of 40 square degrees.}
  \begin{tabular}{c|c|c|c|c}
\hline
  field  & R.A.(J2000),Dec.(J2000) & flux threshold & 
number of sources & brightest source \\
& (h:m:s, d:m:s) & (Jy) &  & (Jy) \\
\hline
1 & (7:57:12,+6:42:00)   & 3  & 786 & 2.51 \\
2 & (5:30:24,-17:36:00)  & 4  & 669 & 3.97 \\
3 & (8:26:48,-10:42:00)  & 4  & 474 & 3.84 \\         
4 & (11:03:36,-14:30:00) & 4  & 745 & 3.08 \\
5 & (8:40:00,-2:30:00)   & 4  & 658 & 3.35 \\
6 & (8:42:24,-8:42:00)   & 5  & 465 & 4.38 \\
7 & (1:05:12,+9:42:00)   & 6  & 743 & 5.07 \\
\hline
\end{tabular}
\label{table:fields40}
\end{table*}

%------------------------------------------------------------
% Figure  : 
%------------------------------------------------------------

\begin{figure*}
\begin{center}
 \includegraphics[width=12cm]{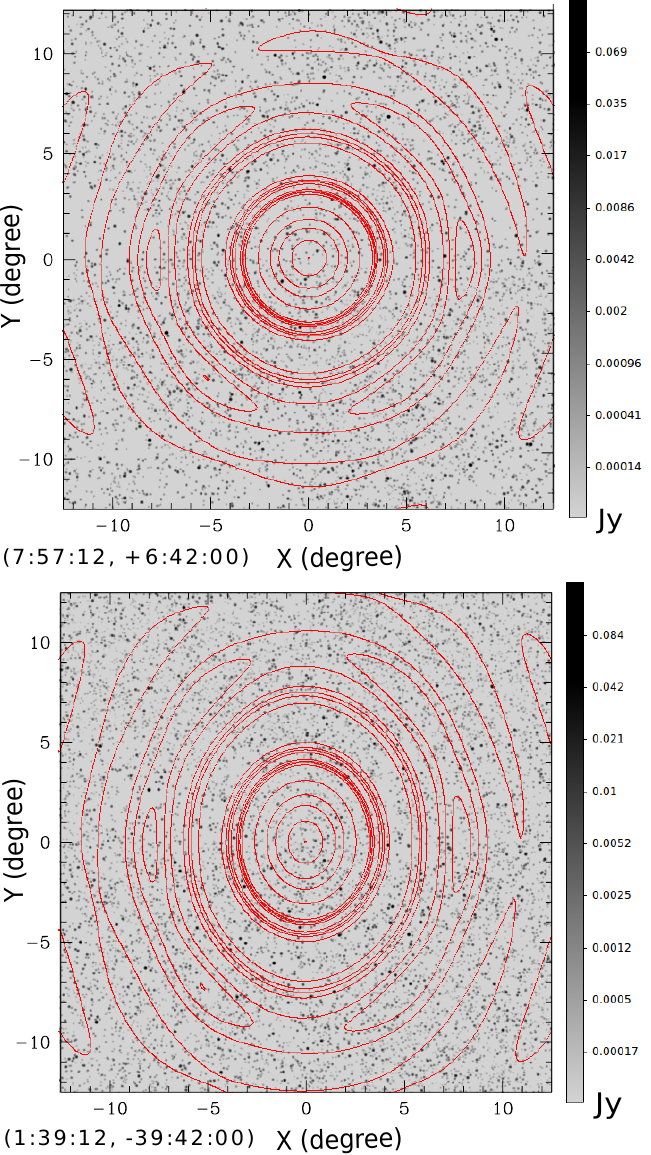}
 \caption{Two examples of the simulated beam patterns 
superposed on the radio maps of two candidate fields at 150 MHz
for 20 (top) and 40 (bottom) square degrees.}
\label{fig:beam-fields}
\end{center}
\end{figure*}

%-----------------------------------------------------------------------------
%-----------------------------------------------------------------------------

Finally we estimate the sidelobe effect for each of the 7+7 candidate fields by convolving the synthesized beam of two SKA1-low stations separated by 5 km (see Figure 1) with the TGSS/GLEAM surveys. The Galactic emission is not included in our current estimates. Figure \ref{fig:beam-fields} shows an example of the simulated beam pattern superposed on the radio source map of the candidate field at 150 MHz for each of the 20- and 40-square-degree samples. The brightest radio sources around the candidate fields are typically a few tens of Jansky (see Table \ref{table:beam1} and Table \ref{table:beam2}), except for two large fields centred at (RA,DEC)=(8:40:00,-2:30:00) and (RA,DEC)=(8:42:24,-8:42:00) around which very bright sources with fluxes of $\sim300$ Jy at 150 MHz have been detected. We convert the radio power contributed by far field sources through sidelobes into the surface brightness temperature over the synthesized primary beam of two SKA1-low stations and present the result in Table \ref{table:beam1} and Table \ref{table:beam2}.  It turns out that for all the seven 20/40-square-degree candidate fields, the sidelobe effects are of the order of magnitudes of $10^{-2}$ mK at 150 MHz - much smaller than the thermal noise ($\sim1$ mK ) required to detect the EoR structures. Therefore, the sidelobe contaminations in our selected samples are luckily negligible for deep imaging of the EoR structures.

%----0.609857-----------------------------------------------------------------
\begin{table*}
%\onecolumn
  \centering
  \caption{Sidelobe effect of the seven candidate fields of 20-square-degrees.}
\begin{tabular}{c|c|c|c|c}
\hline
  field  & R.A.(J2000),Dec.(J2000) & number of sources & brightest source & $T_{\rm beam}$\\
& (h:m:s, d:m:s) & & (Jy) & (mK)\\
\hline
1 &(1:39:12,-39:42:00)  &8753 & 22.31 & $1.35\times10^{-2}$\\
2 &(3:23:36,-60:00:00)  &8320 & 45.93 & $5.88\times10^{-3}$\\
3 &(4:48:00,-13:24:00)  &11324& 24.20 & $1.68\times10^{-2}$\\         
4 &(6:28:24,-49:42:00)  &6223& 46.86 & $1.35\times10^{-2}$\\
5 &(7:00:48,-52:24:00)  &10679& 46.86 & $1.04\times10^{-2}$\\
6 &(7:41:12,-56:24:00)  &8640 & 15.14 & $1.06\times10^{-2}$\\
7 &(23:10:00,-8:06:00)  &9165 & 31.05 & $1.03\times10^{-2}$\\

\hline
\end{tabular}
\label{table:beam1}
\end{table*}

%---------------------------------------------------------------------
% <3,4,5,6 Jy, 40 square degrees
%---------------------------------------------------------------------
\begin{table*}
%\onecolumnop00
  \centering
  \caption{Sidelobe effect of the seven candidate fields of 40-square-degrees.}
  \begin{tabular}{c|c|c|c|c}
\hline
field  & R.A.(J2000),Dec.(J2000) & number of sources & brightest source & $T_{\rm beam}$\\
& (h:m:s, d:m:s) & & (Jy) & (mK)\\
\hline
1 & (7:57:12,+6:42:00)    &10736 &26.39 &$1.88\times10^{-2}$\\
2 & (5:30:24,-17:36:00)   &11088 &47.37 &$2.10\times10^{-2}$\\
3 & (8:26:48,-10:42:00)   &9146 &21.09 &$1.77\times10^{-2}$\\         
4 & (11:03:36,-14:30:00)  &9945 &22.31 &$3.11\times10^{-2}$\\
5 & (8:40:00,-2:30:00)    &8186 &318.82 &$1.83\times10^{-2}$\\
6 & (8:42:24,-8:42:00)    &10067 &318.82&$1.60\times10^{-2}$\\
7 & (1:05:12,+9:42:00)    &10060 &42.77&$2.04\times10^{-2}$\\
\hline
\end{tabular}
\label{table:beam2}
\end{table*}

It is interesting to note that none of the 7 selected fields of 20 square degrees is located within the 7 larger selected fields of 40 square degrees. Namely, the two samples exhibit no overlapped areas in the southern sky. To explore further the phenomenon, we calculate the mean flux and flux variance over the central area of 20-square-degrees for each of the 7 large fields of 40-square-degrees. Similarly, the same estimate is made for a large area of 40-square-degrees around each of the 7 smaller fields of 20-square-degrees. Table \ref{table:40-20sqr} and Table \ref{table:20-40sqr} list the numbers and flux variances of the total radio sources and the brightest ones in the inner (20-square-degree) and outer (40-square-degree) regions for the two samples, respectively. It appears that each of the 7 smaller fields of 20 square degrees exhibits a flux variance of an order of magnitude higher than that of the candidate field itself. Fortunately, the clumpy environment does not bring out extra noise through sidelobes in terms of the above calculations shown in Table 3. By contrast, the flux variance over the central area of 20-square-degrees enclosed in the field of 40-square-degrees for each of the 7 fields is slightly higher than the average value over the whole area, yet exceeds the flux variance over the same area of the small field sample by roughly an order of magnitude.

%%%%%%%%%%%%%%%%%%%%%%%%%%%%%%%%%%%%%%%%%%%%%%%%%%%%%%%%%%%%%%%%%%%%%%%%%%%%
%----0.609857-----------------------------------------------------------------
\begin{table*}
%\onecolumn
  \centering
  \caption{Radio source distribution over a large area of 40-square-degrees 
     around the 20-square-degree candidate fields.}
\begin{tabular}{c|c|c|c|c}
\hline
  field number & R.A.(J2000),Dec.(J2000)& number of sources& brightest source& variance\\
& (h:m:s, d:m:s) & & (Jy) & 40(20) square degrees (Jy)\\
\hline
1 &(1:39:12,-39:42:00) & 409 & 12.90 &0.869 (0.022)\\
2 &(3:23:36,-60:00:00) & 332 & 3.46  &0.149 (0.018)\\
3 &(4:48:00,-13:24:00) & 746 & 9.49  &0.341 (0.009)\\
4 &(6:28:24,-49:42:00) & 452 & 4.82  &0.201 (0.017)\\
5 &(7:00:48,-52:24:00) & 394 & 5.49  &0.242 (0.054)\\
6 &(7:41:12,-56:24:00) & 370 & 3.47  &0.144 (0.026)\\
7 &(23:10:00,-8:06:00) & 664 & 3.95  &0.143 (0.010)\\
\hline
\end{tabular}

\label{table:40-20sqr}
\end{table*}
%----0.609857-----------------------------------------------------------------
\begin{table*}
%\onecolumn
  \centering
  \caption{Radio source distributions within the central area of 20-square-degrees
  for the 40-square-degree candidate fields.}
\begin{tabular}{c|c|c|c|c}
\hline
  field number & R.A.(J2000),Dec.(J2000)& number of sources& brightest source & variance\\
& (h:m:s, d:m:s) & & (Jy) &  20(40) square degrees (Jy) \\
\hline
1 & (7:57:12,+6:42:00)  & 386 &2.51 & 0.119 (0.041)\\
2 & (5:30:24,-17:36:00) & 346 &3.97 & 0.142 (0.073)\\
3 & (8:26:48,-10:42:00) & 250 &3.84 & 0.152 (0.122)\\  
4 & (11:03:36,-14:30:00)& 387 &2.53 & 0.123 (0.099)\\
5 & (8:40:00,-2:30:00) & 345 & 2.45 & 0.066 (0.071)\\
6 & (8:42:24,-8:42:00) &78 & 4.07 & 0.184 (0.355)\\
7 & (1:05:12,+9:42:00) &366 &5.07 & 0.163 (0.060)\\
\hline
\end{tabular}

\label{table:20-40sqr}
\end{table*}

\section{Discussion and Conclusions}\label{sec:discussion}
Using existing catalogs and surveys over a wide frequency band especially in low frequencies, we have selected 7 candidate fields of 20 square degrees in the southern sky, which contain no prominent foreground features (e.g. the Galactic plane, LMC/SMC), no bright diffuse sources (e.g. clusters of galaxies), no bright point sources, and meanwhile exhibit the smallest variance of surface brightness. The sidelobe effect from off-field bright sources is also found to be negligibly small. This work could be regarded as the first step towards preparation of deep imaging of the EoR - the top priority science goal with the forthcoming SKA1-low.

Yet, whether we can achieve the desired sensitivity ($\sim 1$ mK) to detect the EoR structures in the candidate fields depends also on how precisely the sky model can be built and whether all the foreground sources can be subtracted to an acceptable level, among which the extended and diffuse radio sources tend to be the most difficult targets to deal with. There have been no efficient or unique algorithm so far to model the extended sources, despite that many efforts have been made in recent years (e.g. \citealp{Braun13,McKinley15,Trott17}). If some of the candidate fields still contain faint, diffuse and complex radio sources that have not been seen by existing low-frequency surveys, we may meet problems in modeling and removal of these sources, which would in turn increase the noise level of deep imaging. Pre-observations of each of these candidate fields are therefore necessary with some of the SKA pathfinders to further reject the `bad' fields. Recall that not only should the morphology of the extended sources be properly described, but also the spectral index of each component of the extended sources should be known a priori or can be fitted out in the modeling. 
 
We have recently launched a campaign to observe each of the seven selected candidate fields for 100 hours with the MWA. Test observations of two radio-quiet areas with 7.5 degree radii (177 square degrees each) in the southern sky were carried out in 2017, and data reduction and analysis are ongoing. We will make an extensive observational study of these candidate fields in a broad frequency range, towards a deep understanding of the properties of radio sources such as morphologies, spectral indices, structures, clustering, etc. in both the fields and their vicinities. Different imaging algorithms and techniques (weighting, wide-field imaging, sky model construction, CLEAN, etc.) will be applied and further developed to beat down the noises arising from various parameters and environments (thermal noise, confusion noise, deconvolution noise, sidelobe noise, calibration noise, etc.). This will also allow us to estimate the computing capabilities to be required for achieving deep imaging of high dynamical range, and optimize the design and reduce the cost of high performance computing for SKA1-low.

\section*{Acknowledgements}

\addcontentsline{toc}{section}{Acknowledgements}
This work is supported by the National Key R\&D Program of China under grant No. 2018YFA0404601, the Key Projects of Frontier Science of Chinese Academy of Sciences under grant No. QYZDY-SSW-SLH022, and the Strategic Priority Research Program of Chinese Academy of Sciences under grant No. XDB23000000. Q.Z acknowledges the sponsorship from Shanghai Pujiang Program 19PJ1410800 and NSFC of China under grant 11973069. Q.G acknowledges the sponsorship from Shanghai Pujiang Program 19PJ1410700. HYS acknowledges the support from NSFC of China under grant 11973070, the Shanghai Committee of Science and Technology grant No.19ZR1466600 and Key Research Program of Frontier Sciences, CAS, Grant No. ZDBS-LY-7013. S.W.D acknowledges an Australian Government Research Training Programme scholarship administered through Curtin University. We thank Robert Braun for kindly providing us the beam model of SKALA4 used to simulate the beam pattern of the SKA1-low station. Finally we grealy acknowledge the referee, Dr. Emma Chapman, for valuable suggestions which improve the presentaion of this work.

\section{DATA AVAILABILITY STATEMENT}
The data underlying this article will be shared on reasonable request to the corresponding author.  

%This work has made use of data from the Galactic and Extragalactic All-sky MWA survey
%(GLEAM) \citep{Wayth15,Hurleywalker17}, the 150 MHz continuum survey 
%from GMRT (TGSS) \citep{intema16},
%the Haslam 408 MHz All-Sky Map \citep{remazeilles15} and
%the LWA 80 MHz map \citep{dowell17}.



%\bibliographystyle{mnras}
%\bibliography{ref}

%%%%%%%%%%%%%%%%%%%%%%%%%%%%%%%%%%%%%%%%%%%%%%%%%%

% Don't change these lines
\bsp	% typesetting comment
\label{lastpage}
\end{document}